\documentclass[11pt]{article}

\usepackage{fullpage}
\usepackage{latexsym}
\usepackage{amsfonts}
\usepackage{amsmath}

\def\openone{\leavevmode\hbox{\small1\kern-3.8pt\normalsize1}}
\def\RR{{\rm I\kern-.2emR}}
\def\tr{{\rm tr}\; }

\def\cb{{\cal B}}

\def\ca{{\cal A}}

\def\cc{{\cal C}}

\newcommand{\ket}[1]{| #1 \rangle}
\newcommand{\bra}[1]{\langle #1 |}

\newcommand{\inner}[2]{ \langle #1 | #2 \rangle}
\newcommand{\dmelement}[2]{ \langle #1 | #2 | #1 \rangle}
\newcommand{\melement}[3]{ \langle #1 | #2 | #3 \rangle}
\newcommand{\beq}{\begin{equation}}
\newcommand{\eeq}{\end{equation}}
\newcommand{\beqa}{\begin{eqnarray}}
\newcommand{\eeqa}{\end{eqnarray}}

\newtheorem{theorem}{Theorem}
\newtheorem{proposition}{Proposition}
\newtheorem{lemma}{Lemma}

\newtheorem{conjecture}{Conjecture}

\begin{document}

\title{A lower bound on the quantum query complexity of read-once functions}
\author{%
Howard Barnum
\thanks{ This work was done in part while the
author was visiting DIMACS and was supported in part by NSF
under grants EIA 00-80234 and 99-06105.} \\
{\normalsize CCS-3, MS B256, Los Alamos National Laboratory}\\
{\normalsize Los Alamos, NM 87545}\\
{\normalsize\tt barnum@lanl.gov}
\and
Michael Saks\thanks{Research supported by NSF grants 
CCR9988526, EIA 00-80234 and 99-06105.}\\ 
{\normalsize Dept. of Mathematics-Hill Center}\\
{\normalsize 110 Frelinghuysen Road}\\
{\normalsize  Rutgers University}\\
{\normalsize  New Brunswick, NJ}\\
{\normalsize \tt saks@math.rutgers.edu}\\
}
\date{November 12, 2001}

\maketitle

\begin{abstract}
We establish a lower bound of $\Omega{(\sqrt{n})}$ on the 
bounded-error quantum 
query complexity of read-once Boolean
functions, providing evidence for the conjecture
that $\Omega(\sqrt{D(f)})$ is a lower bound for
all Boolean functions.
Our technique extends a result of Ambainis, 
based on the idea that successful computation of a function 
requires
``decoherence'' of initially coherently superposed 
inputs in the query register, having different values of the 
function.
The number of queries is bounded by comparing the
required total amount of decoherence of a judiciously
selected set of input-output
pairs
to an upper bound on the amount
achievable in a single query step.  
We use an extension of this result to
general weights on input
pairs, and  general superpositions of inputs.
\end{abstract}

\section{Introduction and summary of results}
In the {\em quantum query} model of computation,
a query register containing a string $x$  
of $n$ bits is accessed 
by a quantum computer via queries.  In each
query, the computer may ask for a single bit $i$
of the query register, and the value $x_i$ of that bit
is returned; queries are {\em quantum coherent},
which means that a computer may superpose different
query requests $i$ with complex amplitudes $\alpha_i$, 
and is returned a superposition
of the corresponding bit values $x_i$.

%This model is important for several reasons.
%One is that the most spectacular quantum
%algorithms, notably Shor's factoring algorithm \cite{Shor94a}
%\cite{Shor97a}
%are best viewed as applications of a query algorithm,
%e.g. for finding the period of a function by querying
%its table of values (cf. \cite{Cleve99b}).  

The quantum query model is the quantum analog to the
classical boolean decision tree model, and is
at least as powerful as the classical model.
It is of great interest to  compare computation
in these two models, and to see the extent to
which quantum computation gives an advantage over
classical deterministic and randomized computation
in this setting.
One of the major algorithmic results in quantum computation
is Grover's  search algorithm \cite{Grover96a}, which
can be viewed as a
quantum algorithm for computing the $n$-bit OR function
with $O(\sqrt{n})$ queries.  
This compares with the $n$ queries required for
deterministic decision trees and the $\Omega(n)$
queries required by classical randomized trees.
This can be used to speed up brute-force
search for solutions to problems 
(e.g. in NP) with polynomially-checkable solutions.
$\Omega(\sqrt{n})$ is known to 
be a lower bound for OR\cite{Bennett97b} \cite{Grover98b},
perhaps our best piece of evidence that
BQP $\subset$ NP.  

There are two major variants of the quantum query
model: the {\em exact} model and the {\em bounded error} model.
In the exact model, we require that the quantum computation
always output the correct answer, and in the bounded
error model we allow that on any input, the computation
may have a small probability $\epsilon$ of being incorrect.
We write $Q_E(f)$ for the quantum complexity of $f$
in the exact model, and $Q_{\epsilon}(f)$ for the
quantum complexity of $f$ in the bounded error model, 
where $\epsilon$ is the permissible error.  (It is
well known that for $\epsilon \in (0,1/2)$ the value
of $\epsilon$ only affects $Q_{\epsilon}(f)$ within
a constant factor.) We also write $D(f)$ for the determinstic
decision tree complexity of $f$.

In the exact model, there are examples of surprising
speedups, for example the 2 bit XOR can be done
exactly with one quantum query, but there are no
known examples where exact quantum computation provides more
than a constant factor speedup over deterministic
decision trees.  

In the bounded error model, the OR function provides an
example where quantum computation gives a significant
speedup over deterministic (and randomized) decision trees.
In fact, the quadratic speedup for OR is the best speedup
result known for any boolean function.
Perhaps the most important problem in quantum query complexity
is to resolve the following conjecture (which seems to have
been suggested by several researchers):

\begin{conjecture}
For any boolean function $f$ and $\epsilon \in (0,1/2)$, $Q_{\epsilon}(f) = 
\Omega(D(f)^{1/2})$.
\end{conjecture}

The best known result of this type says
that for any $f$ $Q_{\epsilon}(f) = \Omega(D(f)^{1/6})$ (This result
appears in the survey article \cite{BW} and is an improvement
on an earlier $\Omega(D(f)^{1/8})$ bound in \cite{Beals98a},
which is obtained by combining the arguments of \cite{Beals98a}
and an improvement, due to Nisan and Smolensky,
of a result of Nisan and Szegedy\cite{NS92}.)
It should be remarked that
the conjecture is for functions whose domain is all of $\{0,1\}^n$;
for functions whose domain is restricted ({\em promise problems})
there are much better speedups known, see e.g. \cite{Sim}.
In fact, the main component
of Shor's
factoring algorithm \cite{Shor94a}\cite{Shor97a} 
is a query algorithm for  the promise problem of
finding the period of a function by querying its table
of values(cf. \cite{Cleve99b}).  

The main result of this paper is to prove the conjecture for the
class of {\em read-once functions}, those functions expressible
by a boolean formula in which each variable appears at most once.
Our results provide a quantum counterpart to the lower bounds
on the randomized decision  tree complexity of read-once functions
given in \cite{Saks86a} and \cite{Santha91a}.

In \cite{Ambainis2000a}, Ambainis introduced a lower bound
technique for the quantum query model.  He applied this
technique to obtain a $\Omega(\sqrt{n})$
bound for a particular read-once function, the function
which is an OR of $\sqrt{n}$ disjoint ANDs of size $\sqrt{n}$.

%Previously, in the quantum model, $\Omega(\sqrt{n})$ bounds
%were known for some special cases of read-once functions,
%notably the OR itself, and the 
%OR of $\sqrt{n}$ disjoint ANDs, each of size $\sqrt{n}$.

\iffalse
Explicit examples
of functions---such as the uniform binary tree alternating
ANDs and ORs at each level---exist for which a bound of
$\Omega(n^{1/4})$ was
the best known \cite{Beals98a}.  For general Boolean functions,
$\Omega(D(f))^{1/6}$, the sixth root of the classical deterministic
complexity, was the best known bound \cite{Beals98a}.  We conjecture that this can
be sharpened to $\Omega(D(f)^{1/2})$;  our result provides some
modest evidence supporting the conjecture.    
\fi

Our method for obtaining the
$\sqrt{n}$ result
generalizes Ambainis' method; in Section
\ref{sec: general bound} we give a generalization of his technique.
Ambainis'  approach is
based on a thought-experiment in which we imagine the computer
to operate on a {\em superposition} of inputs in the query
register.  The idea is that successful computation of a function 
requires, in the thought-experiment,
``decoherence'' of initially coherently superposed 
inputs in the query register, having different values of the 
function.  This is because successful computation must
correlate input states having different 
values of the function  with nearly orthogonal states 
in the part of the computer where the result is
to be read.  In Ambainis' main results, the inputs are,
essentially, taken to be superposed with equal coefficients.
The number of queries is bounded by comparing the
required total amount of decoherence of a judiciously
selected set of input-output
pairs
to an upper bound on the amount
achievable in a single query step.  Our result generalizes
this technique to give a corresponding result using
the {\em weighted} total decoherence 
of input
pairs (rather than 
just including/excluding pairs via weights
equal to zero or one), 
and  general superpositions of inputs rather than uniform ones.
We anticipate that this result and this approach
will prove useful well beyond the context of read-once
functions to which we apply it in this paper.

\section{Quantum query complexity}

In any quantum computation model, we think of the memory
of the machine as composed of {\em registers}, where
each register has a set of allowed values.  A memory
configuration is an assignment of values to registers.

Each register $R$ is associated to a complex vector
space $H_R$ whose dimension is equal to the number
of allowed values of the register. $H_R$  has a distinguished
orthogonal basis whose members are in one-to-one correspondence
with the possible values of the registers.  We use
the Dirac or ``bra-ket'' notation for complex vector spaces:
elements of such a space are denoted by the notation
$\ket{\phi}$, and  viewed as complex column vectors.
For such a vector $\bra{\phi}$ denotes the dual
row vector whose coordinates are the complex
conjugates of those of $\ket{\phi}$.  
The notation $\inner{\phi}{\psi}$ denotes the
(complex) inner product of $\ket{\phi}$ and
$\ket{\psi}$. The standard basis vectors
of $H_R$ are denoted by $\ket{v}$ where $v$ is an
allowed value for $R$. 

A group of registers can be viewed together
as a single virtual register. 
If $R_1,\ldots,R_k$ are
registers and $R$ is the virtual register obtained
by combining them, then the value set of
$R$ is the product of the value sets of the $R_i$
and the space $H_R$ is naturally isomorphic to the 
tensor product $H_{R_1} \otimes \cdots \otimes H_{R_k}$.
If $v_1,\ldots,v_k$ are possible values 
for $R_1,\ldots,R_k$, then $\ket{v_1,\ldots,v_k}$
is a standard basis element of $H_R$ and
is identified with $\ket{v_1} \otimes \cdots \otimes \ket{v_k}$
which is also written $\ket{v_1}\ket{v_2}\cdots\ket{v_k}$.

In particular, the entire memory can be considered as
a virtual register in this way and is associated
to a complex vector space $H$.  The {\em quantum states}
of this memory are unit vectors in $H$.

Now let's consider the quantum query model.
Here the memory is viewed as divided into
three registers:
the {\em input register} which holds an $n+1$ bit string
$x_0,x_1,\ldots,x_n$ where $x_0$ is fixed to $0$, 
the {\em query
register}
which holds an integer between 0 and $n$,
and the {\em auxiliary memory} which has no restrictions on its
value set.
The query register and auxiliary memory together
comprise the {\em working memory}.
The standard basis of the associated
complex vector space $H$ consists of vectors of the form
$\ket{x,i,z}$ 
in which the input string is $x$, the query is $i$,
and the auxiliary memory is set to $z$.
Thus, a state of the computer is represented  as:

$$
\sum_{x,i,z} \alpha_{x,i,z}\ket{x,i,z},
$$
where for each memory assignment $x,i,z$,
$\alpha_{x,i,z}$ is a complex number, and
$\sum_{x,i,z} |\alpha_{x,i,z}|^2=1$

The space $H$ can be viewed as the tensor
product of two spaces $H_{In} \otimes H_{W}$
where the input space $H_{In}$ is spanned by
the $2^{n}$ basis vectors $\ket{x}$ corresponding
to inputs,
and the work space $H_W$ is spanned
by vectors $\ket{q,z}$ corresponding to
possible contents of the working memory.
The space $H_W$ is further decomposed
as $H_{Q} \otimes H_A$ where
the query space $H_Q$ is spanned by the
$n+1$ query values $\ket{q}$ and the
auxiliary space $H_A$ is spanned by the assignments
$\ket{z}$ to the auxiliary space.  
Thus $\ket{x,i,z}$ is identified with the tensor
products 
$\ket{x} \otimes \ket{i,z}=\ket{x}\ket{i,z}$ 
and $\ket{x}\otimes \ket{i} \otimes \ket{z} = \ket{x}\ket{i}\ket{z}$.

Each computation step is a unitary operator 
on this vector space.  In the query model,
there are two types of operators allowed.
A {\em work space} transformation is one that
operates only on the work space, which means
it is of the form $I_{In} \otimes A$ where
$I_{In}$ is the identity operator on $H_{In}$ and
$A$ acts arbitrarily on $H_W$.
The unitary operator $O$,
called the {\em oracle}, operates as follows:

\begin{eqnarray} \label{oracle action}
O \ket{x,i,z}=
(-1)^{x_i}\ket{x,i,z}\;
\end{eqnarray}

An algorithm is specified by (1) an arbitrary sequence
$U_1,\ldots,U_t$ of work space operators and
(2) a pair of orthogonal projectors $P_0$ on the space $H_W$, i.e.,
a pair of linear maps satisfying $(P_0)^2=P_0$, $P_1^2=P_1$ and
$P_0+P_1=I_W$.  
\iffalse
is an orthogonal decomposition
of $H_W$ 
which means there are associated operators $P_0$ and $P_1$
on $H_W$ where $P_j \ket{v}$ gives the orthogonal
projection of $\ket{v}$
onto $V_j$.
\fi

An algorithm is executed as follows.
The memory is initialized in the basis state with 
the input register set to the input $x$ and
all other registers set to 0.
Then the sequence
$U_1,O,U_2,O,\ldots,U_t,O$ is applied
to the computer.
For $l \in \{1,\ldots,t\}$
the pair $U_l,O$ is called
the $l^{th}$ step of the computation.
Observe that the operations $U_l$ and $O$ leave
the input register unchanged. Formally this means
that the state of the computer is always of the
form
$\ket{x} \otimes \ket{\Psi}$ where $x$ is the input
and $\ket{\Psi}$ is a vector of $H_W$ (generally
not a standard basis state).

The output of the computation is either
0 or 1, determined according to the
following
probability distribution.
If the final state of the computation
is $\ket{x} \otimes \ket{\Psi}$ then the
computation outputs $j$ with
probability equal to $\|P_j\ket{\Psi}\|^2$.
(Note that the definition of $P_0$ and $P_1$
guarantees that the vectors $P_0\ket{\Psi}$
and $P_1\ket{\Psi}$ are orthogonal and
sum to a unit vector, which implies
that the two probabilities sum to 1.)
The process which generates this distribution
is called a
{\em measurement}.

Variants of this model have been considered; in particular
the oracle $O$ can be replaced by a more general transformation
that transforms the work space depending on the value
of the input bit indexed by the query register.
It is well known that this generality can not speed
up the computation by more than a factor of 2.

\iffalse
The computation finishes with an additional unitary transformation,
followed by the measurement of a specified register in the 
standard basis, with the result to be interpreted as the value of 
$f$.  (We
could also omit the final unitary and allow an arbitrary, 
even positive operator-valued, measurement to be made immediately
following the final oracle call.  We would also need to specify
how the measurement result is to be interpreted as a value of $f$.)
\fi

The complexity of the algorithm is measured by the
number of calls $t$ to the oracle.
In the bounded-error quantum query model,  we fix
some $\epsilon<1/2$ and a computation is considered to successfully
compute $f$ if it
$\epsilon$-{\em computes} $f$, which means that for every input,
the probability that the
algorithm gives the wrong answer for that input, is no greater than
$\epsilon$.  
The $\epsilon$-error quantum query
complexity of $f$, denoted $Q_{\epsilon}(f)$ is the minimum
number of steps in an algorithm that $\epsilon$-computes $f$. 
It is well known that the choice of $\epsilon \in (0,1/2)$ only affects
the complexity up to a constant factor.

To avoid confusion, we note that a different way of phrasing
quantum query complexity problems is sometimes used:  the oracle
is said to be a ``black box function,'' $g$.  This $g$ is what
we are calling the ``input,'' (and using the letters
$x$ or $y$ for);  it is {\em not} the function 
$f$ being computed.  The black box terminology sometimes
calls our  $f$ something like $P$, and $P$ is
said to be a {\em property}
of the black box function $g$.  
This model does not represent
the input $g$ 
as a state of a register in the computer.
Rather, the computer is our computer minus the input register,
and a query step 
is an application of the unitary $O_g$ to the computer state.
Generally, $O_g$ is viewed as acting on two registers, an
``input'' register which is homologous to what we have called
the query  register, and an output register.  Its action in
the standard basis is to compute $g$ of the state in the input
register, and write it (in modular arithmetic to ensure unitarity)
in the output register, while keeping the input around.  (An alternative
phase version of the query unitary, similar to (\ref{oracle
action}),  
is sometimes used in this picture,
too.)  We mention this approach primarily to forestall any confusion 
that could arise because the terms ``function''
and ``input'' may used for different things on this approach than on
the one we have adopted.

\section{A general lower bound on quantum query 
complexity of Boolean functions} \label{sec: general bound}

In this section, we present a general extension of Ambainis'
lower bound approach.

Let $f$ be an $n$-variate boolean function whose query
complexity we want to lower bound.
The lower bound is expressed in terms
of a complex vector
$\ket{\alpha}$ of length $2^n$ indexed by inputs (so it is
a member of  $H_{In}$) and a $2^n \times 2^n$ nonnegative
real matrix $\Gamma$ indexed by
pairs of inputs, satisfying $\Gamma_{xy} = 0$ if $f(x)=f(y)$.
For such a matrix $\Gamma$, for each 
$i \in \{1,\ldots, n\}$ we define for $x \in \{0,1\}^n$:

\begin{eqnarray}
\nu_{x,i} &=& \sum_{y: x_i \ne y_i} \Gamma_{xy},
\end{eqnarray}
the total {\em weight} of inputs differing from $x$ on variable $i$.
Further, for $i \in \{1,\ldots,n\}$ and $b \in \{0,1\}$ we define:

\begin{eqnarray*}
\nu_{i}^b&=& \max_{x:f(x)=b} \nu_{x,i}\\
\nu_{i} & =& \nu_{i}^0\nu_{i}^1\\
\nu &=& \max_{j \in \{1,\dots,n\}} \nu_j.
\end{eqnarray*}

The main result of this section is:

\begin{theorem} \label{general theorem}
Let $f$ be an $n$-variate  boolean function.  Let $\ket{\alpha}$
be a nonnegative real valued vector indexed by $\{0,1\}^n$ and $\Gamma$
be a nonnegative real matrix indexed by $\{0,1\}^n \times \{0,1\}^n$
satisfying $\Gamma_{x,y} = 0$ whenever $f(x)=0$ or $f(y)=1$.
If there is a quantum algorithm that 
$\epsilon$-computes $f$ using $t$ queries, then
\begin{equation}
t \ge
\frac{
\bra{\alpha}\Gamma\ket{\alpha}(1 - 2 \sqrt{\epsilon (1 -
\epsilon)})
}{\sqrt{\nu}}
= \Omega \left(\frac{
\bra{\alpha}\Gamma\ket{\alpha}}
{\sqrt{\nu}}
\right)\;.
\end{equation}
\end{theorem}

Buhrman and Szegedy (personal communication) have independently
obtained a similar result.

This should be compared to 
Theorem 6 of
\cite{Ambainis2000a}. In this theorem, Ambainis gives
a lower bound which can be obtained from
the above theorem by letting $\Gamma$ be a 0-1 matrix
and letting $\ket{\alpha}$ be a 0-1 vector.  
Thus  $\ket{\alpha}$ is the characteristic function
of some subset $Z$ of inputs and $\Gamma$ is the characteristic
function of some relation $R$ on $f^{-1}(1) \times f^{-1}(0)$.
(Actually, if we define $X=Z \cap f^{-1}(1)$
and $Y=Z \cap f^{-1}(1)$,
Ambainis' defines the vector $\ket{\alpha}$ to have 
$\alpha_z=1/\sqrt{|X|}$ for $z \in X$ and $1/\sqrt{|Y|}$
for $z \in Y$, but this normalization does not affect the
bounds.)  This choice of $\Gamma$ and $\alpha$ leads
to simplifications of both the numerator and denominator.

When specialized as above, the 
denominator in our expression reduces to the denominator in
Ambainis' theorem 6.
%results.  However, it is entirely analogous to Ambainis' expression 
%$\sqrt{l_{max}}$, except that total $\Gamma$-weight instead of 
%total number is used in defining our $\nu$.
Ambainis defines $l_{x,i}$ to be the number of
$y \in Y$ such that $R(x,y)$ and $x_i \ne y_i$, and $l_{y,i}$ 
similarly.  
%One might expect that if these are large, it reduces the 
%number of queries necessary to compute the function, because it 
%means that a given query index $i$ distinguishes many of the
%$y$'s that must be distinguished from $x$, so the $i$-th
%query has more power.  Of course, there are many $x$'s in $X$,
%and each of them must be distinguished from all the $y$'s 
%such that $R(x,y)$, and similarly for the $y$'s.  
%This is
%why Ambainis' Theorem 2 involves $l$ and $l'$, which are
%the maxima of $l_x,i$ and $l_y,i$ respectively, over $x(y)$ and
%$i$.  This says that no query distinguishes more than at least $l(l')$ 
%relevant  ones (zeroes) from any given zero(one), 
%so it is not surprising that it sets
%an upper bound on the power of each query.
Our parameter $\nu$ specializes to Ambainis' parameter
$l_{max}$ which is defined as the maximum
of the product $l_{x,i} l_{y,i}$, over ones $x$, zeroes $y$, 
and query indices $i$. 
%it turns out to provide a tighter upper
%bound on the power of each query, and a correspondingly 
%tighter lower bound on the number of required queries.  

Ambainis also defines $m (m')$ as the minimum over $x \in X$ 
($y \in Y$) of the number of $y' \in Y$ ($x' \in X$) such
that $R(x,y')$ ($R(y,x')$), which bounds the numerator from below.
% This allows him to bound the 
%numerator below, i.e. bound
%the total decoherence that the algorithm must cause between
%inputs that stand in $R$ to each other.  
Then Ambainis'
Theorem 6
says that the complexity is
$\Omega(\sqrt{mm'/l_{max}})$.
%, (the weaker Theorem 2 has
%$\Omega(\sqrt{mm'/ll'})$).  We do not give an analogous 
%bound for the numerator in our result.  [Mike:  should we try?]

\iffalse
In these theorems, Ambainis provides a bound on the numerator of his
expression, in terms of properties of the relation $R$, 
whereas, although we have a generalization of the expression
he bounds, we have not provided an analogous bound.
Our numerator
involves arbitrary superposition coefficients $\alpha_x$ and 
$\alpha_y$, and arbitrary weights $\Gamma_{xy}$.  Ambainis'
results involve  the special case where
$\Gamma_{xy}$ is the characteristic function 
of a relation $R(x,y)$, where $x$ belongs to a set $X$ of ones of
the function, and $y$ to a set $Y$ of zeroes.  
Ambainis'
restriction is equivalent to restricting $\Gamma_{xy}$
to take only the values $0$ and $1$.  
In other words,
a particular set of matrix elements between inputs is summed.
Ambainis also 
specializes to the case where $\alpha_x = 1/\sqrt{|X|}$,
$\alpha_y = 1/\sqrt{|Y|}$, whereas our result involves 
general coefficients $\alpha_z$.    
(The difference between 
$\alpha$ for inputs $X$ and
$Y$ in Ambainis' case
is due to normalization, and could be omitted without
affecting the lower bounds obtained.)
\fi

In the application to read-once functions in the next section, we will
not need the full generality of Theorem \ref{general theorem}.
In particular, as Ambainis does, we 
restrict $\Gamma$ to be the characteristic
function of a relation.  On the other hand,
the nonuniformity of the
coefficients $\alpha$ will be crucial to our results.

The remainder of this section is devoted to a proof
of Theorem \ref{general theorem}.  The proof is a more or less
straightforward generalization of Ambainis' bound.

Let $U_1,\ldots,U_t$ be a sequence of work space operators
and $P_0,P_1$ be a pair of orthogonal projectors on $H_W$ that specify
an algorithm.
Once we fix an algorithm, then on input $x$, 
the state of the computation
after $j$ steps is of the form $\ket{x}\ket{\Psi_x(j)}$,
where $\ket{\Psi_x(j)} \in H_W$.
(We will normally suppress the index $j$).
Let us consider the set of vectors
$\{\ket{\Psi_x=\Psi_x(t)}:x \in \{0,1\}^n\}$ after $t$
computational steps, but before the final measurement.
%after the $t$ computational steps, but before the final measurement,
%when the input register starts in the states $\ket{x}$.

\begin{proposition}[Ambainis]\label{ambainiscondition}
For a computation to $\epsilon$-compute $f$, it is necessary that
for any $x,y$ such that $f(x)=1$, $f(y)=0$,
\begin{eqnarray}
|\inner{\Psi_x}{\Psi_y}| \le 2 \sqrt{\epsilon(1-\epsilon)}\;.
\end{eqnarray}
\end{proposition}

\begin{proof}
If the computation $\epsilon$-computes $f$, then
%there exist projectors $P_0$, $P_1$ on the nonquery regsisters,
%such that
\begin{eqnarray}
\label{hot diggity}
||P_0 \ket{\Psi_x}||^2 =: \eta_x \le \epsilon, \nonumber \\
||P_1 \ket{\Psi_y}||^2 =: \eta_y \le \epsilon \;.
\end{eqnarray}
Now, 
\begin{eqnarray}
|\inner{\Psi_x}{\Psi_y}| & = &
|\melement{\Psi_x}{P_1}{\Psi_y} + 
\melement{\Psi_x}{P_0}{\Psi_y} |\nonumber \\
& \le & 
|\sqrt{\dmelement{\Psi_x}{P_1}\dmelement{\Psi_y}{P_1}}
+ \sqrt{\dmelement{\Psi_x}{P_0}\dmelement{\Psi_y}{P_0}}| \nonumber \\
& = & \sqrt{(1 - \eta_x)\eta_y}
+ \sqrt{\eta_x (1 - \eta_y)} \le 2 \sqrt{\epsilon (1 - \epsilon)}\;,
\end{eqnarray}
by the Schwarz inequality and $(\ref{hot diggity})$.
\end{proof}

We remark that these necessary conditions are not sufficient
\cite{SBS}.

Define $M$ to be the matrix with elements $M_{xy} = 
|\inner{\Psi_x}{\Psi_y}|$.  (When we want to explicitly
consider the situation at step $l$, we write $M(l)$ for the 
matrix with elements 
$|\inner{\Psi_x(l)}{\Psi_y(l)}|$.)
It is useful to group the inputs according to whether 
$f(x)=0$ or $1$, and view the matrix as a two-by-two 
matrix of block
structure $M_{0,0}$, $M_{0,1}$, $M_{1,0}$, $M_{1,1}$
given by this grouping.   
Proposition \ref{ambainiscondition} involves only the
off-diagonal block, say, $M_{1,0}$.
The general approach of Ambainis involves looking at how
much a single query can decrease the
matrix elements of this off-diagonal block.
Of course, many matrix elements must be considered at
once, because any individual matrix element $M_{xy}$
can be brought
down to zero by a single query to any bit $i$ for 
which $x_i \ne y_i$.  In fact, such a query will reduce
to zero {\em all} $M_{xy}$ such that $x_i \ne y_i$.
However, such a query will 
fail to have any impact on matrix elements for which 
$x_i = y_i$.  There is thus a tradeoff between various
sets of matrix elements. A successful deterministic
classical algorithm must cannily choose $i$'s, depending
on the results of previous queries,  such that
each query distinguishes many inputs that were not 
distinguished by previous ones.  For a probabilistic
classical algorithm, at each query probability may be
distributed between the indices $i$;  and in a 
quantum algorithm, complex {\em amplitude} rather than probability
is distributed over the query indices.  But in each case
there is a tradeoff:  more probability, or more amplitude, 
on a query that distinguishes one set of input pairs, can reduce
the probability, or amplitude, on queries distinguishing
another set.

In order to incorporate such tradeoffs while providing a
necessary condition for $\epsilon$-computation less
complicated than the full set of conditions
implied by Proposition \ref{ambainiscondition}, 
we might consider averaging all the 
off-diagonal-block matrix elements' moduli $|M_{xy}|$.  Since they must
all drop below $\kappa := 2 \sqrt{\epsilon(1 - \epsilon)}$, 
so must their average.  In fact, we
may consider any desired 
positive weighted sum $S$ of the off-diagonal-block
matrix elements, $\sum_{x,y} \mu_{xy} |M_{xy}|$.
For reasons that are still a bit mysterious to us,
it turns out that it is useful to express the weight
$\mu_{xy} = \Gamma_{xy} \alpha_x \alpha_y$, where $\Gamma$
is a nonnegative real matrix and $\ket{\alpha}$ is a unit vector with
nonnegative entries.
On the face of it, this more complex expression
provides no additional generality, but it provides additional
flexibility in the analysis.
The vector $\ket{\alpha}$ can be interpreted, as Ambainis does, 
as
an initial
superposition of inputs in the query register.

As an immediate consequence of Proposition \ref{ambainiscondition}
we have:

\begin{proposition}\label{finalvalue} 
Let $f$ be an $n$-variate boolean function, and let $A$ be
a $t$-step 
quantum query algorithm that attempts to compute $f$.
Let $\ket{\alpha}$ be a unit vector
indexed by $\{0,1\}^n$ with nonnegative real entries
and $\Gamma$ be a matrix indexed by $\{0,1\}^n \times \{0,1\}^n$
with nonnegative real entries 
that satisfies $\Gamma_{x,y}=0$ if $f(x) = f(y)$.
If $A$ $\epsilon$-computes $f$ then:
\begin{eqnarray}
\sum_{xy} \Gamma_{xy} \alpha_x \alpha_y |M(t)_{xy}|
\le \bra{\alpha}\Gamma \ket{\alpha}
\sqrt{2 \epsilon(1 - \epsilon)}\;.
\end{eqnarray}

\end{proposition}

For $l \in \{0,\ldots,t\}$,
let us define
\beqa \label{oop bop shbam}
S_{l}=
\sum_{xy} \Gamma_{xy} \alpha_x \alpha_y|M(l)_{xy}|.
\eeqa
Since $M(0)$ is the all 1 matrix, Proposition \ref{finalvalue}
implies:

\begin{proposition} \label{lowerboundontotal}
For a $t$-step computation to $\epsilon$-compute $f$, 
it must be the case that
\begin{equation}
\label{required decrease}
S_0 - S_t \ge \sum_{xy} \Gamma_{xy} |\alpha_x| |M_{xy}| |\alpha_y| 
(1 - 2\sqrt{\epsilon(1-\epsilon)})\;.
\end{equation}
\end{proposition}

We will now get a lower bound on $t$ by upper bounding
$S_l-S_{l+1}$, the amount that the sum can decrease as the
result of
a single query.
\begin{proposition} \label{prop: decrease in a query}
\beqa
S_l - S_{l+1} \le 2 \sqrt{\nu}\;.
\eeqa
\end{proposition}
The proof of this proposition appears in an Appendix.

If we multiply 
this upper bound on decrease per query by $t$,
this must exceed the difference $S_0-S_t$. This
together  with  Proposition
\ref{lowerboundontotal} completes the proof of the theorem.

\iffalse
We analyze the effect of 
a query, to obtain an upper bound to the amount by which 
each query can decrease the weighted sum.  
To bound the decrease in $S$ per query, 
we use
the fact that the state of the computer before each query
must have unit amplitude distributed among the query
indices.
We know the starting 
value of the sum, and Proposition \ref{ambainiscondition}
gives us an 
upper bound on the final value, if the computation 
$\epsilon$-computes the function.  
We thus have a lower bound on the
decrease in $S$ during a successful computation, and dividing
by an upper bound in the rate of decrease per query
gives a lower bound on the number of queries necessary.
\fi

\section{Read-once Boolean functions} \label{sec: readonce}

A read-once Boolean function is one
which can be written as a formula in propositional
logic, involving each variable $x_i$ (each bit of the input
string) only once. 
Each such function can be represented
by an AND/OR tree.   This is a rooted labeled tree having
$n$ leaves, each corresponding to a different variable (with
some possibly negated),
and where each internal node is labeled either
AND or OR.  Each AND (resp., OR) node in the tree is associated
to a function which
is defined recursively as the
AND (resp.  OR) of the functions computed
by its children.

Without loss of generality, we may assume that all of the
children of an AND node are OR nodes, and vice versa.
Also, we restrict attention to monotone functions,
which are those whose leaves are all nonnegated variables,
since the  query complexity of the function  is preserved
under negation of variables.

\iffalse
be alternately composed either entirely of OR nodes, or entirely
of AND nodes.  We think of the trees as rooted at the top.
Strictly speaking, we also need to include the possibility of 
negating some input variables in order to get all read-once
functions, but obviously the complexity bounds are unaffected
if we restrict ourselves to AND/OR trees 
(or what is the same thing, monotone read-once functions).
\fi

In this section, we use the convention that the
variable
$x$ indicates zeroes of the function, and the
variables $y$ and $z$ indicate ones of the function.

\begin{theorem} \label{read-once bound}
$\Omega(\sqrt{n})$ is a  
lower bound on the bounded-error
quantum query complexity of all
read-once Boolean functions.  
\end{theorem}

We outline the proof technique before providing details.

\iffalse
The proof proceeds by 
induction on the depth of the tree.  
We assume, without loss
of generality that the root is an AND (the
case that it is an OR is essentially the same and is omitted).
Let the root have $r$ children.
We assume that the theorem
holds for each of the functions corresponding
to the $r$ subtrees rooted at children of the root.
\fi
%When it comes time to use this induction step
%in the inductive argument for a $\sqrt{n}$ lower bound on the query
%complexity of all read-once functions,  
%we will need steps of this type alternating with  ORs of functions $g$, 
%but the 
%argument below will cover that type of step too, with obvious minor
%alterations.  

We will apply Theorem \ref{general theorem}.  For this
we need to define the matrix $\Gamma$ and the vector $\ket{\alpha}$.
We identify a subset of $\{0,1\}^n$ called {\em critical} inputs.
These are, intuitively, the inputs on which $f$ is
hardest to compute. 
(The same notion of critical input plays a similar role
in the lower bound proofs for the randomized query complexity
of read-once functions \cite{Saks86a} \cite{Santha91a}).

We also define what it means for
two critical inputs $x \in f^{-1}(0)$ and
$y \in f^{-1}(1)$ to be {\em neighbors}; intuitively
these are pairs of inputs that are hard to distinguish.
We define the matrix $\Gamma$ to be the characteristic
function of the neighbor relation on the set of critical inputs.
Given these choices, 
it will turn out from the definition of critical inputs
that $\sqrt{\nu}$ is always one.

The main work of the proof comes in choosing the
vector $\ket{\alpha}$.  We look for a choice
of the vector $\ket{\alpha}$ that maximizes the
expression in the lower bound of Theorem \ref{general theorem}
(given our particular choice of $\Gamma$).
This (continuous) 
maximization problem is formulated using Lagrange 
multipliers, and gives rise to a set of
first-order conditions.  
We then construct $\ket{\alpha}$ that satisfies the
first order conditions.  

This solution 
is constructed inductively.  Assume that the root
is an AND and has $r$ children and for $i \in \{1,\ldots,r\}$
let $g^i$
denote the function computed at child $i$.
(The case that the root is an OR involves only obvious
minor alterations.)
We write $n_i$ for the number of (boolean) variables in $g_i$.
Thus $n := \sum_{i=1}^r n_i$ is the number 
of (boolean) variables in $f$.  
Assume that 
we have determined $\ket{\alpha^i}$ 
for each of the $g_i$. We construct $\ket{\alpha}$
in terms of these.  
Further we show that if
$\ket{\alpha^i}$ gives a bound of 
$\kappa \sqrt{n_i}$ for each of the $g_i$, then
$\ket{\alpha}$ gives a bound of $\kappa \sqrt{n}$ for $f$.

\iffalse
We assume the coefficients $w_x$, $w_y$ to be chosen
so as to make the lower bound of theorem 1 (given the 
assumptions on $\Gamma$) maximal, not only for $f$, but for
each of the $g_i$.

Using the known bound of
$\kappa \sqrt{r}$ for OR or AND as our base case, this inductively
implies the bound of $\sqrt{n}$ for any AND/OR tree.
In establishing the induction step, we will 
make extensive use of the
first-order conditions for Lagrange multiplier optimization of
Theorem 1's lower bound expression (as simplified by our additional
assumptions).
\fi

We proceed with the detailed proof.
We have the read-once function $f$ represented by tree $T$,
and express $f$ as $g^1 \wedge \ldots \wedge g^r$
where $g^i$ are the functions computed at  
the children of the root labeled by AND.

In choosing a $\Gamma$ and $\ket{\alpha}$ for applying Theorem
\ref{general theorem},
we will focus our attention on
{\em critical inputs}.  An input is critical
if for each AND node, at most one child evaluates
to 0 and for each OR node, at most one child evaluates to 1.
A critical input in $f^{-1}(1)$ is a {\em critical one}
and a critical input in $f^{-1}(0)$ is a {\em critical zero}.

\iffalse
These inputs are defined recursively.  For example, the unique 
critical one of an AND is an input which is all ones; 
the $r$ critical ones of an OR of $r$ variables each have 
all zeroes except for a single variable, which is 1.  A critical zero
of an AND is all ones except for a single zero; a critical zero of an
OR is all zeros.  A critical one (zero) of an AND/OR tree is defined 
(recursively) as an assignment of values to its leaves such that
when any subtree is replaced with a leaf having the 
subtree's value under that assignment, the resulting tree is still 
a critical one (zero).

A bit $i$ is said to be {\em critical for an input x}
if flipping the bit $x_i$ changes $f(x)$ from zero
to one.  
Note that despite the temptation, critical inputs 
may not be characterized merely by the existence of 
critical bits for them, nor may it be said that all ones
in a critical one are critical for it.  It may be shown, 
however, that the critical bits in a critical one (zero)
are ones (zeroes), and that flipping a critical bit 
in a critical one (zero) results in a critical zero (one).
\fi

We write $X$ (resp. $Y$) for the set of critical zeros (resp.,
critical ones)
of $f$.  For $i \in \{1,\ldots,r\}$
we write $X^i$ (resp., $Y^i$) for the set of critical zeros
(resp., critical ones) of $g^i$.
We use the letter $x$ to denote an element of $X$,
the letters $y$ and $z$ to denote elements of $Y$.
Also $x^i$ denotes an element of $X^i$ and $y^i$ and $z^i$
denote elements of $Y^i$.

Observe that since the root is an AND,
a critical one $y$ may be written in the form $y=y^1\ldots y^r$,
where for each $j$, $y^j$ is a critical one of $g^j$.
For a critical zero $x$, exactly one of the children
of the root evaluates to 0.  We say that $x$ is of
{\em type $i$}, for $i \in \{1,\ldots,r\}$, 
if $g^i$ evaluates to 0.  A critical zero $x$ of type $i$
may be written in the form
$x=z^1 \ldots z^{i-1}x^i z^{i+1}\ldots z^r$,
where $x^i$ is a critical zero of $g_i$ and
for $j \neq i$, $z^j$ is a critical one of $g^j$. 

Let $x \in X$ and $y \in Y$ and let $i$ be the type
of $x$.  We say that $y=y^1 \ldots y^r$ and 
$x=z^1 \ldots z^{i-1}x^i z^{i+1}\ldots z^r$,
are {\em neighbors} provided that $z^j=y^j$ for $j \neq i$
and $x^i$ and $y^i$ are {\em neighbors} (defined recursively).
We denote by $R$ the neighbor relation on $X \times Y$
and $R_i$ the neighbor relation on $X_i \times Y_i$.
It is easy to see that two neighbors differ on exactly
one input variable and consequently, for any 
critical input $w$ and any $j \in \{1,\ldots,n\}$,
$w$ has at most one neighbor that differs from
it on variable $j$.

When we apply Theorem \ref{general theorem} we take 
$\Gamma$ to be the characteristic function of
the relation $R$.
It is easily seen that for this $\Gamma$, the parameter
$\nu$ appearing in the denominator in Theorem \ref{general theorem}
is just 1:  by the last sentence of the previous
paragraph,
the quantity $\nu_{w,j}$ is at most 1 for any critical input $w$
and $j \in \{1,\ldots,n\}$.

%From now on, our variables will take only critical inputs as 
%values.  
\iffalse
We use a notation in which 
concatentation of variables represents
a string which is the concatenation of the values of the variables,
rather than the product of their values. 

Since $f$ is an AND, critical
zeroes of $f$ 
are such that each of the $g$'s but one (the $i$-th) takes the value
one, so they have the form:
\begin{eqnarray}
x = y^1 y^2...y^{i-1} x^i y^{i+1}...y^r 
\end{eqnarray}

for some $i$, while the ones ($y$'s) of the function
have the form: 
\begin{eqnarray}
y = y^1 y^1 ...y^r.
\end{eqnarray}

For an $x$ (critical one of $f$), we call $i$ the {\em type}
of the input $x$.
\fi

Having fixed $\Gamma$, we now want to choose $\ket{\alpha}$.
Without loss of generality we
will take the coordinates of $\ket{\alpha}$ to
be nonnegative real numbers and assume that they
are zero outside of $X \cup Y$.
We look for an $\ket{\alpha}$ such that the
lower bound expression in Theorem \ref{general theorem}
is maximum.      This means we want to solve:

\begin{eqnarray}
\max \sum_{(x,y) \in R} \alpha_x \alpha_y \nonumber \\
{\rm S.T.}~~\sum_x \alpha_x^2 + \sum_y \alpha_y^2 = 1
\end{eqnarray}

From Lagrange multiplier optimization for the above problem,
we get the first-order conditions (FOCs):
\begin{eqnarray}
\alpha_x = \cc \sum_{y: (x,y) \in R} \alpha_y\;, \nonumber \\
\label{FOC}
\alpha_y = \cc \sum_{x: (x,y) \in R} \alpha_x\;.
\end{eqnarray}

Here $\cc$ is a constant to be determined.

Suppose we find $\cc$ and a unit vector $\ket{\alpha}$ 
satisfying (\ref{FOC}).  If we multiply the first
FOC by $\alpha_x$ and sum on $x \in X$ we get
that the objective function is equal to 
$\frac{1}{\cc} \sum_{x \in X} \alpha_x^2$.  Similarly
if we multiply the second FOC by $\alpha_y$ and sum on $y \in Y$
we have that the objective function equals
$\frac{1}{\cc} \sum_{y \in Y} \alpha_y^2$.  This implies
that $\sum_{x \in X} \alpha_x^2 = \sum_{y \in Y} \alpha_y^2 = \frac{1}{2}$
and the value of the objective function is $\frac{1}{2\cc}$.
We will prove:

\begin{lemma}
There is a nonnegative real unit vector $\ket{\alpha}$
satisfying (\ref{FOC}) with $\cc = 1/\sqrt{n}$.
\end{lemma}
Theorem \ref{read-once bound} now follows immediately from the lemma
and Theorem \ref{general theorem}

The proof of the lemma is by induction.  For the base case, we take $f$ to
be the univariate functions $f(x_1)=x_1$.  For this
function $X=\{0\}$, $Y=\{1\}$ and $\alpha_0=\alpha_1=\frac{1}{\sqrt{2}}$
solves the FOC with $\cc=1$.
For the induction step, we assume that the lemma holds
for each of the functions $g_i$ and prove that it holds
for $f$.  

Let $x=y^1\ldots y^{i-1}x^iy^{i+1}\ldots y^r$ be an element of $X$.
All neighbors $y$ of $x$ must have a critical one $y^i$
in the $i$-th place that is a neighbor of $x^i$,
while agreeing with $x$ in the 
other places, and thus be of the form:
\begin{equation}
y^1 \ldots y^i \ldots y^r\;.
\end{equation}
So,
\begin{equation} \label{lagrangefocsx}
 \alpha_x \equiv \alpha_{y^1...x^i...y^r} = \cc \sum_{y^i:
(x^i,y^i) \in R_i} \alpha_{y^1..y^i...y^r}.
\end{equation}
Similarly, for $y=y^1 \ldots y^r \in Y$,
\begin{eqnarray}\label{lagrangefocsy}
\alpha_y \equiv \alpha_{y^1....y^r} = \cc \sum_i \sum_{x^i: (x^i,y^i) \in R_i}
\alpha_{y^1 \ldots y^{i-1} x^i y^{i+1} \ldots y^r}\;.
\end{eqnarray}

By the induction hypothesis, 
for each $i \in \{1,\ldots,r\}$, we have a unit vector
$\alpha^i$  that satisfies the
first-order conditions
for $g^i$:
\begin{eqnarray} \label{lowerlevelfocsy}
\alpha^i_{y^i} = \cc_i \sum_{x^i: R_i(x^i,y^i)} \alpha^i_{x^i}\;, \\
\label{lowerlevelfocsx}
\alpha^i_{x^i} = \cc_i \sum_{y^i: R_i(x^i,y^i)} \alpha^i_{y^i}\;.
\end{eqnarray}
with $\cc_i=\frac{1}{\sqrt{n}}$.
 
We proceed to establish the induction step.  We guess that
the weights at the top level are the product of the weights at the
next level down, up to a constant which can depend on the {\em type}
of the input whose weight we are computing.  (Here, only the $x$'s
have distinct types, depending on which of the $g_i$ has the value 
zero.)  
\begin{eqnarray} \label{ansatz1}
\alpha_y \equiv \ca \alpha_{y^1 \ldots y^r} = 
\alpha^1_{y^1} \alpha^2_{y^2} \ldots \alpha^r_{y^r} \\
\label{ansatz2}
\alpha_x \equiv \alpha_{y^1 \ldots x^i \ldots y^r} = \cb_i  
\alpha^1_{y^1} \alpha^2_{y^2} \ldots \alpha^i_{x^i} \ldots \alpha^r_{y^r} \;.
\end{eqnarray}

We check these guesses by plugging them into both sides of 
(\ref{lagrangefocsx})
and (\ref{lagrangefocsy}), respectively, obtaining:
\begin{equation} \label{wx}
(\alpha_x = ) \cb_i \alpha_{y^1...x^i...y^r} = \cc \ca 
\alpha^1_{y^1} \ldots \alpha^{i-1}_{y^{i-1}}...
\left(\sum_{y^i: (x^i,y^i) \in R_i} {\alpha^i_{y^i}}
\right)\alpha^{i+1}_{y^{i+1}}
\ldots \alpha^r_{y^r}\;,
\end{equation}
\begin{equation} \label{wy}
(\alpha_y = ) \ca \alpha^1_{y^1} \ldots \alpha^r_{y^r}
= \cc \sum_i \cb_i \alpha^1_{y^1}...\alpha^{i-1}_{y^{i-1}}
\left( \sum_{x^i: R_i(x^i, y^i)} \alpha^i_{x^i} \right)
\alpha^{i+1}_{y^{i+1}} \ldots \alpha^r_{y^r}\;.  
\end{equation}

The parenthesized sums
in these expressions evaluate to 
$\alpha^i_{x^i}/ \cc_i$ and $\alpha^i_{y^i}/\cc_i$, 
respectively, from the lower level Lagrange FOCs
(\ref{lowerlevelfocsy}-\ref{lowerlevelfocsx}).
So our guess solves the
higher level FOCs.
In the equation (\ref{wx}) for $w_x$, this requires
\begin{equation}\label{zoot suit}
\cb_i = \frac{\cc \ca}{\cc_i}\;.
\end{equation}
In the equation (\ref{wy}) for $w_y$, we obtain:
\begin{eqnarray}
w_y = \cc \sum_i \cb_i \frac{1}{\cc_i} w_{y^1}...w_{y^i}...w_{y^r}\;,
\end{eqnarray}
and (since the weight product on the RHS is $i$-independent), 
this requires (substituting for $\cb_i$ using (\ref{zoot suit}))
\begin{equation}\label{constant C}
\ca = \cc \sum_i \frac{\cc \ca}{\cc_i^2}\;,
~~{\rm i.e.,}~~\frac{1}{\cc^2} = \sum_i \frac{1}{\cc_i^2} \;.
\end{equation}

Since $\cc_i=1/\sqrt{n_i}$ we deduce 
$\cc=1/\sqrt{\sum_i n_i}=1/\sqrt{n}$ as required.

\appendix
\section{Proof of Proposition \ref{prop: decrease in a query}}

%This follows immediately from Proposition \ref{ambainiscondition}.
Without loss of generality for our purposes, we may take
$\Gamma$ to be symmetric, upper triangular, or lower
triangular.  
\iffalse
The initial value of the weighted sum will be 
\begin{equation}
\sum_{xy} \Gamma_{xy} |\alpha_x| |M_{xy}| |\alpha_y|\;.
\end{equation}
Thus the sum must decrease by at least the amount given
in the following proposition:
\fi
% We plan to do two things: first, allow arbitrary positive
% weighted sums of matrix elements with weights $\Gamma_{xy}$.
% Second, analyze the weights in part, at least, as the
% coefficients of normalized vectors.
% Bounding the change in the sum of all matrix elements that 
% can be accomplished with $t$ queries is often too loose a 
% procedure.  Let's say we have a certain bound on the number
We proceed to establish an upper bound on the magnitude of
the decrease of the weighted sum (\ref{oop bop shbam}) 
in a single query.
It will be convenient to assume $\Gamma$ is symmetric.
Define 
$\ket{\Psi^i_x}$ as the component $\ket{\Psi_x}$ having
$i$ in the query register, so that

\begin{equation}
\label{keen}
\ket{\Psi_x} = \sum_i \ket{\Psi_x^i}\;.
\end{equation}

Then
\begin{equation}
M = \sum_i M^i
\end{equation}
where $M^i$ is defined via $(M^i)_{xy} = 
\inner{\Psi^i_x}{\Psi^i_y}$.  (Just write
$M_{xy}$ as an inner product, use (\ref{keen}), 
and note that the $i\ne j$ terms are zero.)
To reduce clutter, define $\rho^i_{xy} := 
\alpha_x^* M^i_{xy} \alpha_y$.  
\iffalse
(The notation $\rho$
is chosen because, if the input register is prepared
in the superposition $\sum_x \alpha_x \ket{x} + \sum_y \alpha_y
\ket{y}$, $\rho^i$ is the ``relative density matrix'' of
the query register, relative to the state $i$ of the 
query register.  That is, it is the total computer density
matrix, projected onto the subspace with $i$ in the
query register, and then partial-traced over everything
but the input register.  $\rho^i$ are not normalized, but
their traces sum to one; indeed
$\sum_i \rho^i$ is the reduced density matrix of the
input register.)  
\fi
In this notation we want to upper bound
$\sum_i\sum_{xy}\Gamma_{xy}|\rho^i_{xy}|$.  We consider the inner sum first.
We first upper bound $|\rho^i_{x,y}|$ as a linear combination
of $\rho^i_{xx}$ and $\rho^i_{yy}$.  For this purpose
we introduce a nonnegative matrix $\beta=\beta_{xy}$, whose
entries we will specify later.  We have:
\begin{equation}
|\rho^i_{xy}| \le \sqrt{\rho^i_{xx} \rho^i_{yy}} = 
\sqrt{\rho^i_{xx} \frac{1}{\beta_{xy}} \beta_{xy} \rho^i_{yy}}
\le \frac{1}{2} (\beta_{xy} \rho^i_{xx} + \frac{1}{\beta_{xy}}
\rho^i_{yy})
\;.
\end{equation}
(The first inequality is due to the positivity of $\rho^i$, which
requires that the determinant of any principal minor be positive;
the second is the arithmetic-geometric mean inequality).  
Then 
\begin{eqnarray} \label{loopdeloo}
\sum_{xy: x_i \ne y_i} \Gamma_{xy} 
|\rho^i_{xy}| \nonumber \\
\le \sum_{x}\sum_{y: y_i \ne x_i} \Gamma_{xy}
\frac{1}{2} (\beta^i_{xy} \rho^i_{xx} + \frac{1}{\beta^i_{xy}}
\rho^i_{yy}) = 
\sum_{x}\sum_{y: y_i \ne x_i} \Gamma_{xy}
\beta^i_{xy} \rho^i_{xx}
\;.
\end{eqnarray}

The last equality is just due to the symmetry under $x \leftrightarrow y$.
%(Let us not forget that $\Gamma_{xy}=0$ when $f(x) \ne f(y)$.)

\iffalse
Now let us define for every input $x$ and variable $i$
\begin{equation}
\nu_{x,i} := \sum_{y: f(x) \ne f(y) \& x_i \ne y_i} \Gamma_{xy}\;,
\end{equation}

the total {\em weight} of inputs differing from $x$ on variable $i$.
(This is an analogue of Ambainis' $l_{x,i}$, the total {\em number} 
of inputs $y$ such that $R(x,y)$ and $x_i \ne y_i$.)
\fi
We now define
\begin{equation}
\beta^i_{xy} = \sqrt{\frac{\nu_{y,i}}{\nu_{x,i}}}\;,
\end{equation}
where $\nu_{x,i}$ was defined at the beginning of this section.

The last expression in (\ref{loopdeloo}) becomes:
\begin{equation}
\sum_x \rho_{xx}^i \sum_{y: y_i \ne x_i}
\Gamma_{xy} \sqrt{\frac{\nu_{yi}}{\nu_{xi}}}\;.
\end{equation}

Recalling the definition of $\nu_i^{b}$ at the beginning
of the section, we can bound this
expression by:
\iffalse
To go further, we assume a uniform (in $y$) upper bound $\nu_i$ 
on the $\nu_{yi}$ that appear in the above sum (in order to factor it
out of the inner sum); such a bound is
$\max_{y: y_i \ne x_i} \nu_{yi}\;.$  
This gives us
\fi
\begin{eqnarray*}
\sum_x \rho_{xx}^i \sum_{y: y_i \ne x_i}
\Gamma_{xy} \sqrt{\frac{\nu_{i}^{1-x_i}}{\nu_{xi}}}
& \le & \sum_x \rho_{xx}^i \sqrt{\nu_{xi} \nu_i^{1-x_i} \ }\\
& \le & \sum_x \rho_{xx}^i \sqrt{\nu_{i}^0 \nu_i^{1} \ }\\
& = & \sum_x \tr \rho^i \sqrt{\nu_i \ }\\
& \le & \sum_x \tr \rho^i \sqrt{\nu \ }\\
\end{eqnarray*}
\iffalse
To factor everything but $\rho_{xx}^i$ out of the $x$ sum, we use 
the same bound again:
\begin{equation}
\le \sum_x \rho_{xx}^i \sqrt{\max_{x}[
\nu_{xi} \max_{y: y_i \ne x_i} \nu_{yi}]}
= \tr \rho^i \sqrt{\max_{x}[
\nu_{xi} \max_{y: y_i \ne x_i} \nu_{yi}]}. 
\end{equation}
\fi
Summing on $i$ then yields:
\begin{equation} \label{dum dee dum}
\sum_i \sum_{xy: x_i \ne y_i} \Gamma_{xy} |\rho^i_{xy}|
%\le \sum_i \tr \rho^i  \max_j \sqrt{\max_{x} [\nu_{xj}
%\max_{y:y_j \ne x_j} \nu_{yj}]}
\le \sqrt{\nu} 
\end{equation}

\iffalse
We name the square of the quantity on the RHS $\nu$.
Note that $\nu = \max_{j, x y : x_j \ne y_j} \nu_{xj} \nu_{yj}$.
\fi

This equation is an important lemma, which we 
use
to establish our bound on the decrease of the weighted
sum in a single query.  For quantities which change in 
a query, we distinguish the post-query quantity by priming
it.
\begin{eqnarray}
S'-S & := &
\sum_{xy} \Gamma_{xy} ( |\rho_{xy}| - |\rho'_{xy}|) = 
\sum_{xy} \Gamma_{xy} ( |\rho_{xy} - \rho'_{xy}|)
\nonumber \\
& \le & 
\sum_i \sum_{xy} \Gamma_{xy} ( |\rho^i_{xy} - {\rho'}^i_{xy}|)
\end{eqnarray}

We can bound the term in parentheses by noting that since $\rho^i$ is
the input register density matrix relative to $i$ in the query index
the query multiplies density matrix elements $\rho^i_{xy}$ by 
a factor
$(-1)^{x_i y_i}$, leaving them unchanged if $x_i = y_i$.  
%(We may 
%be able to make the same argument for the most general form of queries.)
Thus
we obtain
\begin{eqnarray}
\sum_i \sum_{xy: x_i \ne y_i}
\Gamma_{xy} ( |\rho^i_{xy} - {\rho'}^{i}_{xy}|)
\le 
\sum_i \sum_{xy: x_i \ne y_i}
\Gamma_{xy} |\rho^i_{xy}| +  |{\rho'}^i_{xy}|.
\end{eqnarray}

We can then apply Eq. (\ref{dum dee dum}) to each of these terms,
obtaining:
\begin{equation}
S - S' \le \sqrt{\nu} \sum_i (\tr \rho + 
\tr \rho') = 2 \sqrt{\nu}\;.
\end{equation}

%%%%%%%%%%%%%%%%% BIBLIOGRAPHY IN THE LaTeX file !!!!! %%%%%%%%%%%%%%%%%%%%%%%%
%% This is nothing else than the IEEEsample.bbl file that you would         
%%
%% obtain with BibTeX: you do not need to send around the *.bbl file        
%%
%%---------------------------------------------------------------------------%%
%
%
%%---------------------------------------------------------------------------%%


\begin{thebibliography}{10}

\bibitem{Grover96a}
Lov Grover,
\newblock ``A fast quantum mechanical algorithm for database search,''
\newblock {\em Proceedings of the 28th Annual ACM Symposium on the Theory of
  Computing (STOC)}, pp. 212--219, May 1998.

\bibitem{Bennett97b}
C.~H. Bennett, G.~Brassard, E.~Bernstein, and U.~Vazirani,
\newblock ``Strengths and weaknesses of quantum computing,''
\newblock {\em SIAM Journal on Computing}, vol. 26, pp. 1510--1523, 1997.

\bibitem{Grover98b}
L.~Grover,
\newblock ``How fast can a quantum computer search?,''
\newblock 1998,
\newblock arXiv.org e-print quant-ph/9809029.

\bibitem{BW}
H.~Buhrman and R.~de~Wolf,
\newblock ``Complexity measures and decision tree complexity,''
\newblock {\em Theoretical Computer Science},
\newblock to appear. (Available at {\tt http://www.cwi.nl/~rdewolf/}).

\bibitem{Beals98a}
Robert Beals, Harry Buhrman, Richard Cleve, Michele Mosca, and Ronald de~Wolf,
\newblock ``Quantum lower bounds by polynomials,''
\newblock {\em FOCS '98}, pp. 352--361, 1998.

\bibitem{NS92}
N.~Nisan and M.~Szegedy,
\newblock ``On the degree of boolean functions as real polynomials,''
\newblock {\em Computational Complexity}, pp. 301--313, 1994.

\bibitem{Sim}
D.~Simon,
\newblock ``On the power of quantum computation,''
\newblock {\em SIAM J. Comp.}, vol. 26, pp. 1474--1483, 1997.

\bibitem{Shor94a}
P.~W. Shor,
\newblock ``Algorithms for quantum computation: discrete logarithms and
  factoring,''
\newblock {\em Proc. 37th ann. symp. on the foundations of computer science},
  pp. 56--65, 1994.

\bibitem{Shor97a}
P.~W. Shor,
\newblock ``Polynomial-time algorithms for prime factorization and discrete
  logarithms on a quantum computer,''
\newblock {\em SIAM J. Comp.}, pp. 1484--1509, 1997.

\bibitem{Cleve99b}
R.~Cleve,
\newblock ``The query-complexity of order-finding,''
\newblock 1999,
\newblock arXiv.org e-print quant-ph/9911124.

\bibitem{Saks86a}
M.~Saks and A.~Wigderson,
\newblock ``Probabilistic {B}oolean decision trees and the complexity of
  evaluating game trees,''
\newblock {\em Proceedings of the 27th {IEEE} Symposium on the Foundations of
  Computer Science ({FOCS})}, pp. 29--38, 1986.

\bibitem{Santha91a}
M.~Santha,
\newblock ``On the {M}onte-{C}arlo {B}oolean decision tree complexity of
  read-once formulae,''
\newblock {\em Proceedings of the 6th {IEEE} Structure in Complexity Theory},
  pp. 180--187, 1991.

\bibitem{Ambainis2000a}
A.~Ambainis,
\newblock ``Quantum lower bounds by quantum arguments,''
\newblock {\em Proceedings of the 32nd Annual {ACM} Symposium on the Theory of
  Computing (STOC)}, pp. 636--643, 2000.

\bibitem{SBS}
M.~Szegedy, H.~Barnum and M.~Saks
\newblock ``Quantum decision trees and semidefinite programming,''
\newblock submitted, 2001.

\end{thebibliography}
\end{document}